\documentclass[12pt,twoside]{article}
\usepackage{fleqn,espcrc1}

\usepackage{graphicx}
\usepackage{epsfig}
\usepackage[figuresright]{rotating}

\newcommand{\AmS}{{\protect\the\textfont2
  A\kern-.1667em\lower.5ex\hbox{M}\kern-.125emS}}

\title{$N^*$, $\Lambda^*$, $\Sigma^*$ and
$\Xi^*$ Resonances from $J/\Psi$ and $\Psi'$ Decays }

\author{Bing-Song Zou\address{CCAST (World Laboratory),
        P.O.~Box 8730, Beijing 100080
        and Institute of High Energy Physics, CAS,
        P.~O.~Box 918(4), Beijing 100039, P.R.China}}
       
\begin{document}

\maketitle

\begin{abstract}
Vector charmonium states $J/\Psi$ and $\Psi'$ are copiously produced by  
$e^+e^-$ annihilations at the Beijing Electron Positron Collider (BEPC).
Their decays provide an excellent place for studying excited nucleons and
hyperons -- $N^*$, $\Lambda^*$, $\Sigma^*$ and $\Xi^*$ resonances. The
systematic study of these excited nucleons and hyperons can provide us
with critical insights into the internal quark structure of baryons.

\end{abstract}

\vskip 5mm

The Institute of High Energy Physics at Beijing runs an electron-positron
collider (BEPC) with a general purpose solenoidal detector, 
the Beijing Spectrometer (BES)\cite{BES}, 
which is designed to study exclusive final states in $e^+e^-$
annihilations at the center of mass energy from 3000 to 5600 MeV.
In this energy range, the largest cross sections are at the $J/\Psi(3097)$
and $\Psi'(3686)$ resonant peaks. At present, the BES has collected about
30 million $J/\Psi$ events and 3.7 million $\Psi'$ events. More data are
going to be taken. Here I want to show you that their decays provide an
excellent place for studying excited nucleons and hyperons -- $N^*$,
$\Lambda^*$, $\Sigma^*$ and $\Xi^*$ resonances. 

\begin{figure}[htbp]
\vspace{-1.6cm}
\hspace{1.5cm}\includegraphics[width=14cm,height=6cm]{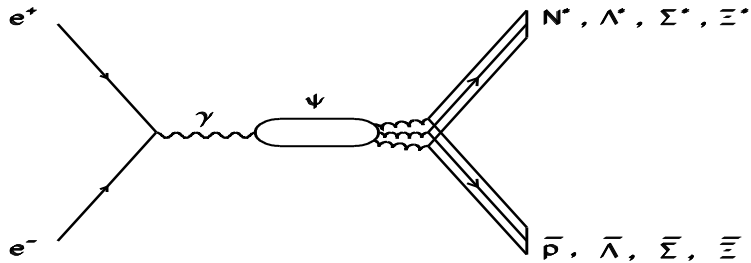}
\vspace{-1.8cm}\caption{Feynman graphs for $\bar pN^*$,
$\bar\Lambda\Lambda^*$, $\bar\Sigma\Sigma^*$ and $\bar\Xi\Xi^*$ production
from $e^+e^-$ collision through $J/\Psi$ or $\Psi'$ meson.}
\label{fig:1}
\end{figure}

\vspace{-0.4cm}
The corresponding Feynman graphs for the production of these excited
nucleons and hyperons are shown in Fig.~\ref{fig:1} where $\Psi$
represents either $J/\Psi$ or $\Psi'$.
Since the vector charmonium $\Psi$ decays through three gluons and gluons
are flavor blind, the strange s quarks are produced at the same level as
the non-strange u and d quarks.  Table~\ref{table:1} lists some interested
$J/\Psi$ decay branching ratios\cite{PDG}. The $p\bar p$,
$\Lambda\bar\Lambda$, $\Sigma^0\bar\Sigma^0$ and $\Xi\bar\Xi$ are indeed 
produced at similar branching ratios. The branching ratios for 
$\bar pN^*$, $\bar\Lambda\Lambda^*$, $\bar\Sigma\Sigma^*$ and
$\bar\Xi\Xi^*$ are expected to be of the same order of magnitude if one
ignores the phase space effect.
The $\bar\Omega\Omega^*$ channels have thresholds above or very close to
the mass of $\Psi'$ and cannot be studied here.

\vspace{-0.3cm}
\begin{table}[htb]
\caption{ $J/\Psi$ decay branching ratios (BR$\times 10^3$) for some
interested channels\cite{PDG}}
\label{table:1}
\renewcommand{\arraystretch}{1.2} 
\begin{tabular}{ccccccc}
\hline
$p\bar p$ & $\Lambda\bar\Lambda$ & $\Sigma^0\bar\Sigma^0$ & $\Xi\bar\Xi$
& $\Lambda\bar\Sigma^-\pi^+$ & $pK^-\bar\Lambda$ & $pK^-\bar\Sigma^0$\\
\hline
$2.14\pm 0.10$ & $1.35\pm 0.14$ & $1.27\pm 0.17$ & $1.8\pm 0.4$ &
$1.1\pm 0.1$ & $0.9\pm 0.2$ & $0.29\pm 0.08$\\
\hline
$p\bar n\pi^-$  & $p\bar p\pi^0$  &  $p\bar p\pi^+\pi^-$ &
$p\bar p\eta$ & $p\bar p\eta'$ & $p\bar p\omega$ &
$K^-\Lambda\bar\Xi^+$ ?\\
\hline
$2.0\pm 0.1$ & $1.1\pm 0.1$ & $6.0\pm 0.5$ & $2.1\pm 0.2$ & 
$0.9\pm 0.4$ & $1.3\pm 0.3$ & $K^+\bar\Lambda\Xi^-$ ? \\
\hline
\end{tabular}\\
\end{table}

\vspace{-0.3cm}

All channels listed in Table~\ref{table:1} are relative easy to be
reconstructed by BES. For example, for $K^-\Lambda\bar\Xi^+$, we can
select events containing $K^-$ and $\Lambda$ with $\Lambda\to p\pi^-$,
then from missing mass spectrum of $K^-\Lambda$ we should easily identify
the very narrow $\bar\Xi^+$ peak. The $K^-\Lambda\bar\Xi^+$ channel is a
very good place for studying $\Xi^*\to K\Lambda$. At present, not much is
known about $\Xi^*$ resonances\cite{PDG}. Only the ground $\Xi(1318)$
state and the first excitation state $\Xi^*(1530)$ are well established. 
There has not been a single new piece of data on $\Xi^*$ resonances since
PDG's 1988 edition. Various theoretical predictions by rather different
physical pictures\cite{Capstick,Glozman} are not challenged due to the
lack of data. With $J/\Psi$ and $\Psi'$ experiments at BEPC and upgraded
BEPCII in near future, we expect to complete the $\Xi^*$ resonance
spectrum as well as the $N^*$, $\Lambda^*$ and $\Sigma^*$ resonance
spectra. 

Among three-body channels listed in Table~\ref{table:1},
$\Lambda\bar\Sigma^-\pi^+$ and $pK^-\bar\Lambda$ can be used to study
$\Lambda^*\to\Sigma\pi$ and $NK$; $\Lambda\bar\Sigma^-\pi^+$
and $pK^-\bar\Sigma^0$ can be used to study
$\Sigma^*\to\Lambda\pi$ and $NK$; Channels containing p can be
used to study $N^*\to K\Lambda$, $K\Sigma$, $N\pi$, $N\pi\pi$, $N\eta$,
$N\eta'$ and $N\omega$. Many other channels not listed in
Table~\ref{table:1} can also be used to study these baryon
resonances.

In fact, the Feynman graphs in Fig.~\ref{fig:1} are almost identical to
those describing the $N^*$ electro-production process if the direction of
the time axis is rotated by $90^o$. The only difference is that the
virtual photon here is time-like instead of space-like and couples to
$NN^*$ through a real vector charmonium meson $\Psi$. So all $N^*$
decay channels which are presently under investigation at
CEBAF(JLab, USA)\cite{Burkert}, ELSA(Bonn,Germany)\cite{Klempt}, 
GRAAL(Grenoble, France) and Spring8(KEK, Janpan) with   
real photon or space-like virtual photon can also
be studied at BEPC complementally with the time-like virtual photon.
In addition, for $\Psi\to\bar NN\pi$ and $\bar NN\pi\pi$, the $\pi N$
and $\pi\pi N$ systems are limited to be pure isospin $1/2$ due to isospin
conservation. This is a big advantage in studying $N^*$ resonances from
$\Psi$ decays, compared with $\pi N$ and $\gamma N$ experiments which
suffer difficulty on the isospin decomposition of $1/2$ and 
$3/2$\cite{Workman}. 

\begin{figure}[htb] 
\begin{minipage}[t]{50mm}
\centerline{\epsfig{file=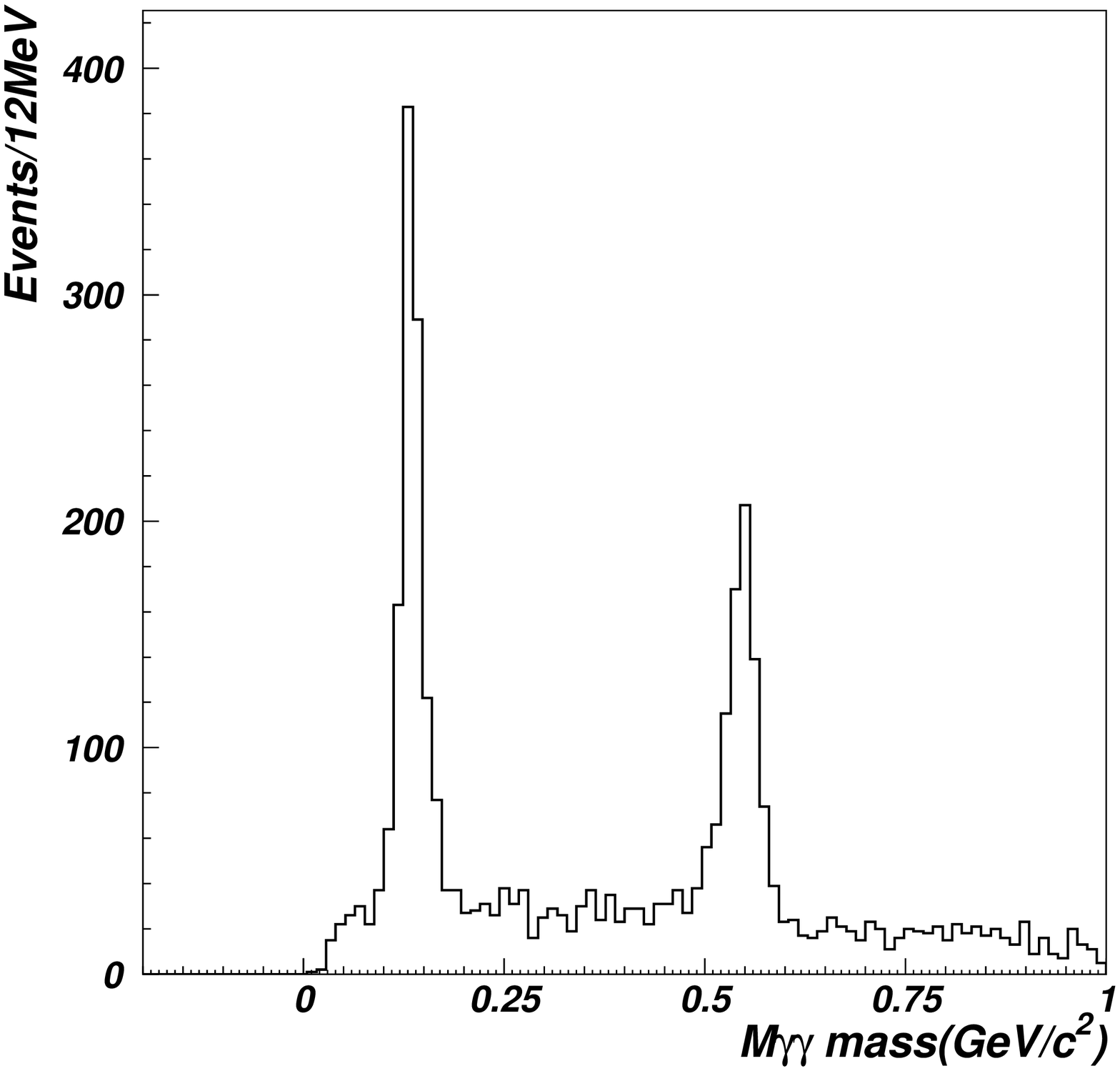,height=2.in,width=2.in}}
\vspace{-1.cm}\caption[]{$\gamma \gamma$ invariant mass spectrum after 4C
fit for $J/\Psi\to\bar pp\gamma\gamma$}
\label{fig:2}
\end{minipage}
\hspace{\fill}
\begin{minipage}[t]{50mm}
\centerline{\epsfig{file=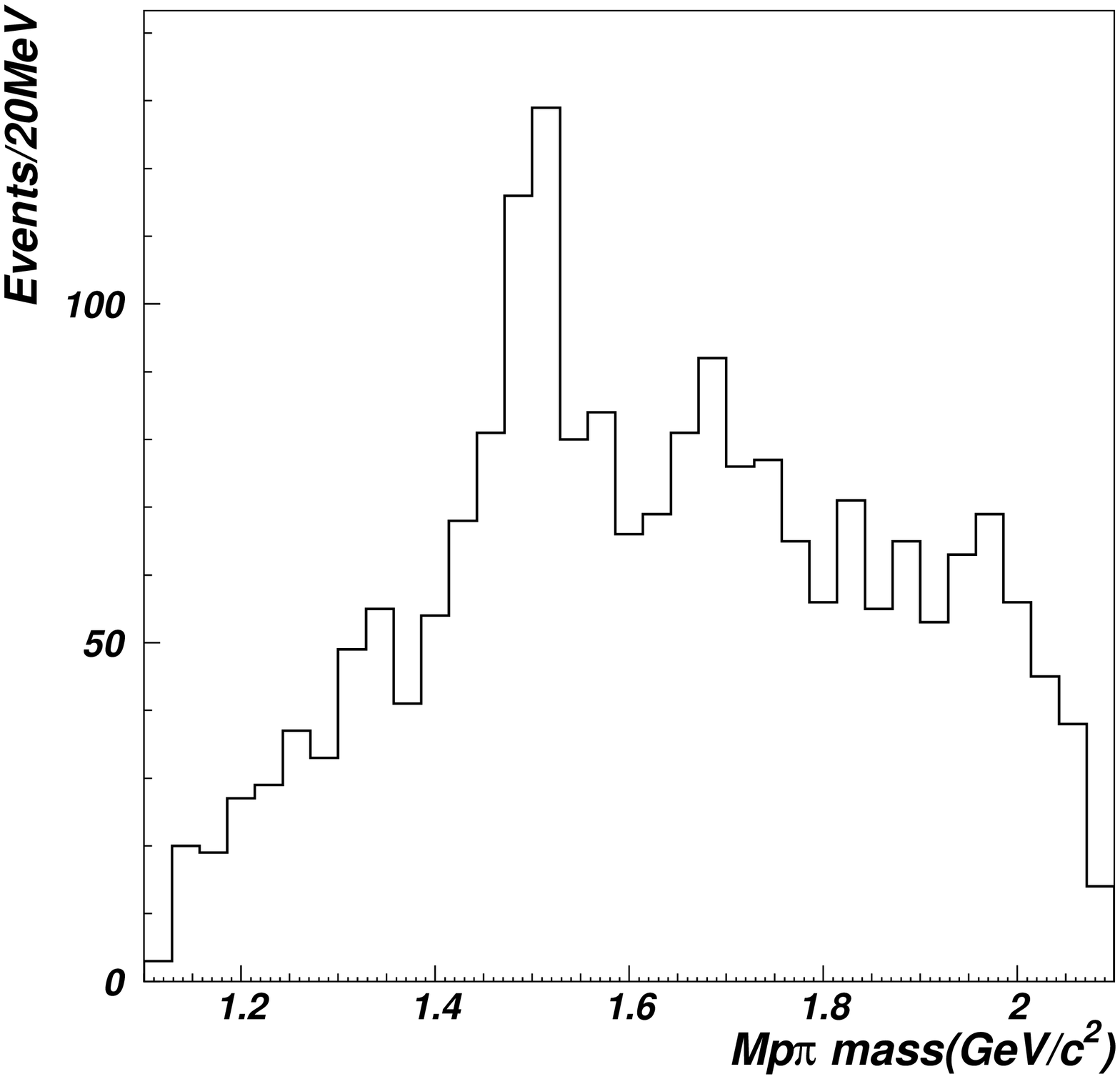,height=2.in,width=2.in}}
\vspace{-1.cm}\caption[]{$p\pi^0$ invariant mass spectrum for
$J/\Psi\to\bar pp\pi^0$.}
\label{fig:3}
\end{minipage}  
\hspace{\fill}
\begin{minipage}[t]{45mm}
\centerline{\epsfig{file=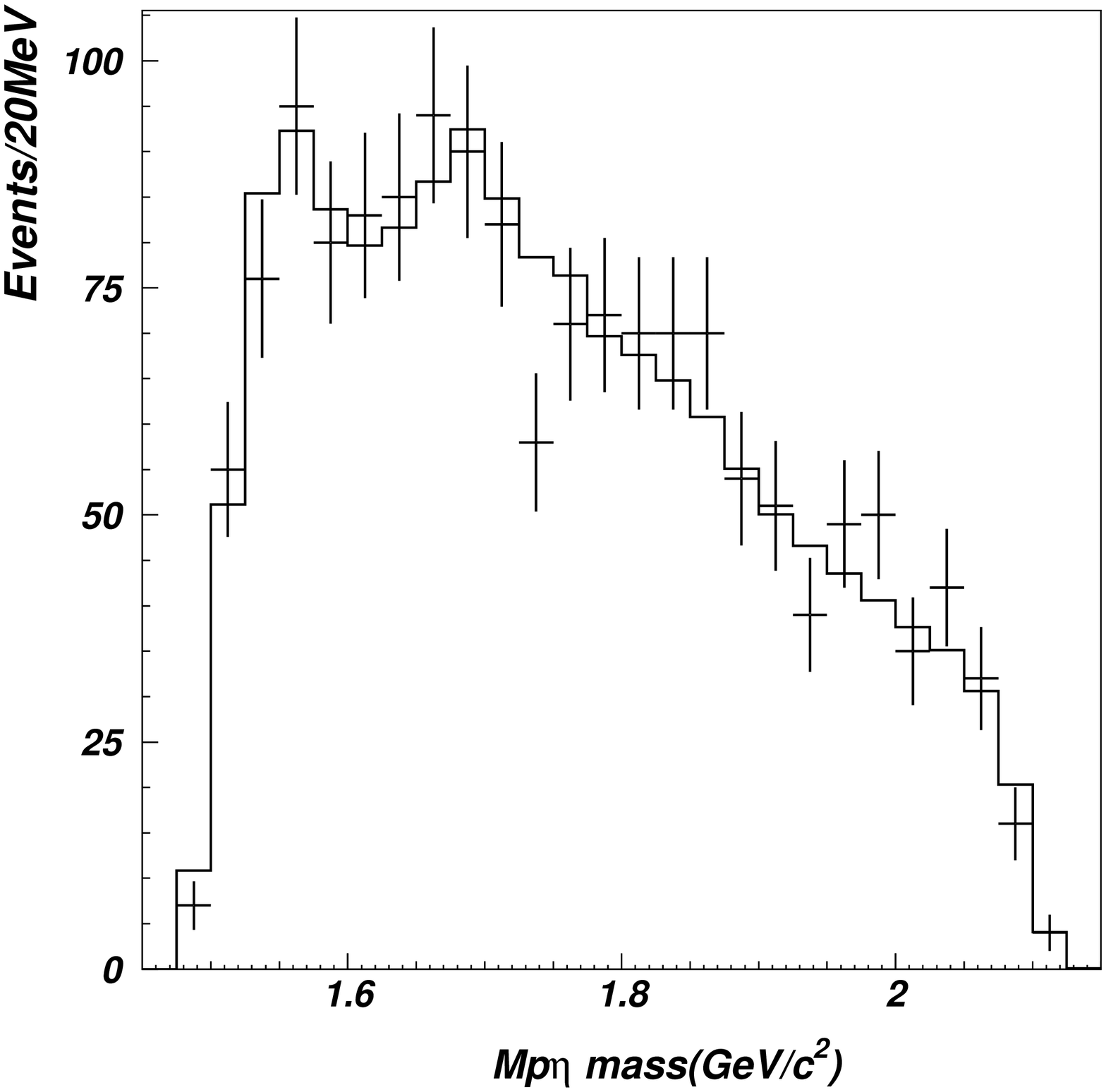,height=2.in,width=2.in}}
\vspace{-1.cm}\caption{$p\eta$ invariant mass spectrum for $J/\Psi\to\bar
pp\eta$.}
\label{fig:4}
\end{minipage}  
\end{figure}

Based on 7.8 million $J/\Psi$ events collected at BEPC before 1996,
the events for $J/\Psi\to\bar pp\pi^0$ and $\bar pp\eta$ have been
selected and reconstructed with $\pi^0$ and $\eta$ detected in their
$\gamma\gamma$ decay mode\cite{Lihb}. For selected $J/\Psi\to\bar
pp\gamma\gamma$ events, the invariant mass spectrum of the $2\gamma$ is
shown in Fig.\ref{fig:2}. The $\pi^0$ and $\eta$ signals are clearly
there. The $p\pi^0$ invariant mass spectrum for $J/\Psi\to\bar
pp\pi^0$ is shown in Fig.~\ref{fig:3} with clear peaks around 1480 and 
1650 MeV. The $p\eta$ invariant mass spectrum for
$J/\Psi\to\bar pp\eta$ is shown in Fig.~\ref{fig:4} with clear enhancement
around the $p\eta$ threshold, peaks at 1540 and 1650 MeV; both have 
been determined to have $J^{P}={1\over 2}^-$ by a PWA analysis\cite{Lihb}.
With more than 20 million new $J/\Psi$ data collected in last few months,
many more channels are expected to be analyzed soon.

On theoretical side, the coupling of $\Psi\to\bar pN^*$ 
provides a new way to probe the internal quark-gluon structure of the
$N^*$ resonances\cite{Zou1}. In the simple
three-quark picture of baryons, as shown in Fig.~\ref{fig:1}, three
quark-antiquark pairs are created independently via a symmetric
three-gluon intermediate state with no extra interaction other than the
recombination process in the final state to form baryons. This is quite
different from the mechanism underlying the $N^*$ production from the
$\gamma p$ process where the photon couples to only one quark and
unsymmetric configuration of quarks is favored.  Therefore the processes
$\Psi\to\bar pN^*$ and $\gamma p\to N^*$ should probe different aspects of
the quark distributions inside baryons.
Since the $\Psi$ decay is a glue-rich process, it is also regarded as a
good place for looking for hybrid $N^*$\cite{Page}.

In summary, the $J/\Psi$ and $\Psi'$ experiments at BEPC provide an
excellent place for studying excited nucleons and hyperons -- $N^*$,
$\Lambda^*$, $\Sigma^*$ and $\Xi^*$ resonances. 
Since baryons represent the simplest system in which the three
colors of QCD neutralise into colorless objects and the essential
non-Abelian character of QCD is manifest, the systematic
study of these baryons will provide us with critical insights into the
nature of QCD in the confinement domain\cite{Klempt,Isgur}.

\end{document}